# Lightweight Quantum Security Protocols

Tzonelih Hwang[1*], Yen-Jie Chen[2], Chia-Wei Tsai[3] and Cheng-Ching Kuo[4]

[1,2,4]*Department of Computer Science and Information Engineering, National Cheng Kung University, No. 1, University Rd., Tainan City, 70101, Taiwan, R.O.C.*

[3]*Department of Computer Science and Information Engineering, National Taitung University, No.369, Sec. 2, University Rd., Taitung City, Taiwan, R.O.C.*

[1] hwangtl@csie.ncku.edu.tw (corresponding author)

[2] mathmatic83@gmail.com

[3] cwtsai676@stust.edu.tw

[4] p76071072@mail.ncku.edu.tw

## Abstract

Inspired by the semi-quantum protocols, this paper defines the lightweight quantum security protocols, in which lightweight participants can only operate two out of four very lightweight quantum operations. Subsequently, this study proposes a Lightweight Mediated Quantum Key Distribution (LMQKD) protocol as an example to disclose the feasibility and advantage of the lightweight quantum protocol. In the proposed protocol, a dishonest third party (TP) with complete quantum capabilities helps two lightweight quantum users establish a secure key. The lightweight quantum users are allowed to perform only: (1) unitary operations and (2) reflecting qubits without disturbance. The proposed protocol has been showed to be robust under the collective attack.

**Keywords:** Quantum cryptography, Semi-quantum key distribution, untrusted third-party.

# 1. Introduction

Since the first quantum key distribution protocol－BB84 was published by Bennett et al. [1] in 1984, various quantum cryptographic protocols have been proposed [2-14]. Most of these protocols allow the participants to have complicate quantum capabilities such as generating entangled states, possessing quantum memory and so on. However, these quantum devices are still expensive and difficult to implement to date. As a result, these protocols are not very practical.

To enable "classical" participants to perform quantum protocols, Boyer et al. [15,16] proposed the concept of "Semi-Quantum" protocol, in which two kinds of users, the quantum users (or servers) and the classical users are defined. The quantum users usually have complete quantum capabilities, i.e., he/she can generate quantum states (e.g., single photon and entangled quantum states), perform measurement (e.g., X-basis, Z-basis, or Bell measurement), store qubits in quantum memory, and so on, whereas the classical users are allowed to perform only three out of four deliberately simplified operations, called semi-quantum operations here: (1) measuring qubits in Z basis $\{|0\rangle, |1\rangle\}$ (2) preparing qubits in Z basis (3) reordering qubits via a delay line, and (4) reflecting qubits without any disturbance. Based on these semi-quantum operations, two kinds of semi-quantum key distribution (SQKD) protocols, namely Measure-Resend and Randomization-Based SQKD protocols, were proposed in [15, 16]. Since then, various semi-quantum protocols and security analyses [15-25] have been proposed.

The above-mentioned SQKD protocols can only let a quantum user and a classical user to share a secret key. In 2015, the mediated SQKD (MSQKD) protocol, particularly allowing two classical users to share a secret key with the help of a dishonest quantum third party (TP), was proposed [26]. Later, Liu et al. [27] proposed a mediated SQKD protocol without invoking quantum measurement for the classical users in 2018.

Among these four above mentioned semi-quantum operations, photon reordering operation appears to be quite complicate in implementation [28]. This observation motivates this paper to research into lightweight quantum protocols. To justify the definition of "lightweight" protocols, it is better to list the "lightweight" quantum operations from the most lightweight to the less ones as follows [29-30]: (1) reflecting qubits without disturbance, (2) performing single-qubit unitary operations, (3) measuring qubits in Z-basis, (4) generating qubits in Z-basis. Based on these lightweight quantum operations, lightweight quantum protocols can be defined as follows.

**Definition**. **Lightweight Quantum Protocol**

A quantum protocol is called lightweight if the lightweight participants only perform two out of the above four mentioned lightweight quantum operations to complete the protocol.

It should be noted that BB84 can be called lightweight based on this definition because the lightweight participants only perform the lightweight quantum operations i.e. Alice performs (2) and (4) and Bob performs (2) and (3). However, BB84 requires two users equip with different quantum operations to distribute a key. To extend the applications of lightweight quantum protocols, an interesting question to ask is: Can we allow two lightweight quantum users with the same lightweight quantum operations to share a key. To provide a solution, this work proposes a lightweight MQKD as an example to demonstrate the feasibility of the lightweight quantum protocols defined above, in which a dishonest TP helps the two lightweight participants to distribute the secret keys.

In particular, this work tries to take the MQKD protocol as an example to show how to design a lightweight MQKD (LMQKD), in which the participants are only with two lightweight quantum operations (i.e., (1) reflecting the qubits without disturbance, (2) performing unitary operations). Security analyses are provided to prove that the proposed lightweight MQKD is robust under the collective attack.

It is very interesting to note that the lightweight quantum operations so selected by the participants in the proposed LMQKD are without quantum measurement operation. From a different point of view, it is like that the participants deliberately delegate the quantum measurement operations to a server, TP, which is exactly the central idea in the research of Measurement-Device-Independent (MDI) protocols [36-53]. Though not all MDI protocols are lightweight protocols, there are do some MDI protocols which are lightweighted. For example, Xu [44] proposed an MDI protocol, which is similar to the proposed LMQKD, from the realizational point of view, that also supports the idea of lightweight quantum security protocols. However, the main contribution of the paper is to define the new lightweight quantum protocols, which we believe is worth of mentioning in consideration of practical implementation of quantum security protocols.

Take [54] as the other example. If the lightweight participants are selecting the (2) and (3) lightweight quantum operations, then the protocol can automatically do without subjecting to the trojan horse attack because the original qubits do not have to be reflected out to the other participants by the receiver.

The concept of lightweight quantum security protocol is not only for the design of QKD protocols. It can also be extended to the design of other quantum security protocols, such as quantum authenticated QKD protocols, quantum secret sharing protocols, quantum private comparison protocols so on and so forth.

The rest of the paper is organized as follows. Section 2 describes the proposed protocol. The security analyses are described in Section 3. The comparison of our protocol and the existing three-party QKD protocols is provided in Section 4. Finally, conclusion remarks are given in Section.5.

## 2. Proposed scheme

This section first introduces the unitary operations $\sigma_Z$, $\sigma_X$, $H$, used in our proposed protocol:

$$H = \frac{1}{\sqrt{2}}(|0\rangle\langle 0| + |0\rangle\langle 1| + |1\rangle\langle 0| - |1\rangle\langle 1|) = \frac{1}{\sqrt{2}}\begin{bmatrix} 1 & 1 \\ 1 & -1 \end{bmatrix}$$

$$\sigma_z = |0\rangle\langle 0| - |1\rangle\langle 1| = \begin{bmatrix} 1 & 0 \\ 0 & -1 \end{bmatrix} \qquad \text{Eq. (1)}$$

$$\sigma_x = |0\rangle\langle 1| + |1\rangle\langle 0| = \begin{bmatrix} 0 & 1 \\ 1 & 0 \end{bmatrix}$$

One can use these unitary operations to transfer the state of a qubit. For example, one can apply the $\sigma_X$ operation on $|0\rangle$ and measure the qubit with Z-basis to get the $|1\rangle$ as the measurement result. The state transitions of single photons by these unitary operations are shown in the Table 1.

**Table 1** The state transitions of single photons

| State \ Unitary operation | $|0\rangle$ | $|1\rangle$ | $|+\rangle$ | $|-\rangle$ |
|---|---|---|---|---|
| $\sigma_z$ | $|0\rangle$ | $-|1\rangle$ | $|-\rangle$ | $|+\rangle$ |
| $\sigma_x$ | $|1\rangle$ | $|0\rangle$ | $|+\rangle$ | $|-\rangle$ |
| $H$ | $|+\rangle$ | $|-\rangle$ | $|0\rangle$ | $|1\rangle$ |

Based on the Eq. (1) and Table 1, we can further perform the unitary operations on a Bell state, and the transition of the Bell state is described as follows.

Let $|\varphi^+\rangle = \frac{1}{\sqrt{2}}(|00\rangle+|11\rangle)$ denote a Bell state. If each qubit in this Bell state is performed with one of the unitary operations $\sigma_z$, $\sigma_x$ or $H$, respectively. The states changed and the combinations of unitary operations are summarized in the **Figure 1**. For example, if the first qubit is performed with $\sigma_z$ and the second qubits is performed with $\sigma_z$, then the Bell state will still be $|\varphi^+\rangle$. If the first qubit of the Bell state is performed with $\sigma_z$ and the second qubit of the Bell state is performed with $H$, then the Bell Measurement of this Bell state will be either $|\varphi^+\rangle$ or $|\psi^-\rangle$.

| Initial state | Operation performed on the first qubit | Operation performed on the second qubit | Changed state |
|---|---|---|---|
| $|\phi^+\rangle$ | Operation $\sigma_Z$ | Operation $\sigma_Z$ | $|\phi^+\rangle$ |
| | Operation $\sigma_Z$ | Operation $\sigma_X$ | $|\psi^-\rangle$ |
| | Operation $\sigma_Z$ | Operation $H$ | $|\psi^-\rangle, |\phi^+\rangle$ |
| | Operation $\sigma_X$ | Operation $\sigma_Z$ | $|\psi^-\rangle$ |
| | Operation $\sigma_X$ | Operation $\sigma_X$ | $|\phi^+\rangle$ |
| | Operation $\sigma_X$ | Operation $H$ | $|\psi^-\rangle, |\phi^+\rangle$ |
| | Operation $\sigma_H$ | Operation $\sigma_Z$ | $|\psi^-\rangle, |\phi^+\rangle$ |
| | Operation $\sigma_H$ | Operation $\sigma_X$ | $|\psi^-\rangle, |\phi^+\rangle$ |
| | Operation $\sigma_H$ | Operation $H$ | $|\phi^+\rangle$ |

**Figure 1**. State transition of the Bell state after performing unitary operations

This study uses the relationship shown in Figure 1 to propose a lightweight MQKD, which requires the following assumptions:

- There are ideal quantum channels (i.e. noiseless and non-lossy) connected between TP and each participant, respectively.
- There are public classical channels connected between TP and each participant, respectively.
- There is an authenticated channel shared between two participants.
- The two participants can only perform (1) the above mentioned unitary operations,

and (2) reflecting photons.

- The untrusted TP can perform any quantum operations and any attacks.

The procedure of the proposed protocol is illustrated in the following (see also **Figure 2**).

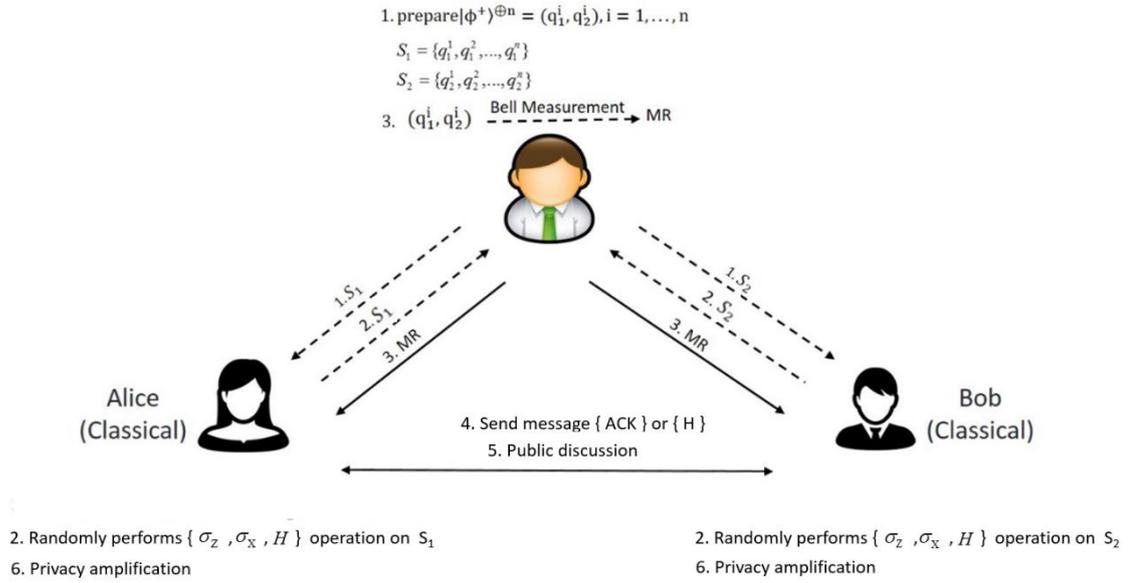

**Figure 2**. The proposed LMQKD

**Step.1** TP generates $n$ Bell states $|\phi^+\rangle$, and sends all the first and all the second qubits of the Bell states, $q_1^i$, $q_2^i$, $1 \leq i \leq n$, one-by-one respectively to Alice and Bob.

**Step.2** Upon receiving a qubit, $q_1^i$ ($q_2^i$), from TP, Alice (Bob) performs a unitary operation randomly chosen from $\{\sigma_Z, \sigma_X, H\}$ on the qubit to form $q_1^i{}'$ ($q_2^i{}'$) and then sends it back to TP, $1 \leq i \leq n$.

**Step.3** While TP receives the two qubits respectively from Alice and Bob, TP will perform a Bell Measurement on the qubit pair. After measuring all these $n$ qubit pairs one-by-one in order, TP will obtain $n$ Bell Measurement results **MR**={ $MR_1, \ldots, MR_n$ }, and send **MR** to Alice and Bob.

**Step.4** Alice and Bob first confirm receiving of the same **MR**. Alice and Bob will then disclose to each other the unitary operations performed on the qubit pairs respectively in order via an authenticated classical channel. If the unitary operation performed on a qubit is chosen from $\{\sigma_Z, \sigma_X\}$, then the message is denoted as "ACK". Otherwise if the unitary operation is $H$, then it is "H". For a particular pair of qubits, there will be four possible cases {ACK, H}, {H, ACK}, {ACK, ACK}, {H, H}.

**Step.5** Based on the messages of the unitary operations performed on the qubit pairs, two groups can be identified in all qubit pairs.

**Group 1. With $H$ operation**

This group includes three cases {ACK, H}, {H, ACK}, {H, H}. For all the qubit pairs in this group, Alice and Bob have to reveal the exact operation $\sigma_z$ or $\sigma_x$ in the ACK to each other. With this information, Alice (Bob) can detect the existence of an eavesdropper (including a malicious TP). For example, for the $ith$ qubit pair, if the first qubit is performed with $\sigma_z$ and the second qubit is performed with $H$, then according to **Figure** 1, the $MR_i$ should be either $|\phi^+\rangle$ or $|\psi^-\rangle$. If the corresponding $MR_i$ does not satisfy the state transition diagram shown in **Figure** 1, Alice and Bob will terminate the protocol and start from the beginning.

**Group 2. Without $H$ operation**

This group only includes the case {ACK, ACK}. In this case, Alice (Bob) can obtain raw key bits by distinguishing the exact unitary operation in the ACK. For example, for a Bell Measurement result $|\phi^+\rangle$, if Alice performed $\sigma_z$ on the first qubit of this Bell state, then by **Figure** 1, she knows that Bob also performed $\sigma_z$ on the second qubit of this Bell state. Similarly, for a Bell Measurement result $|\psi^-\rangle$, if Bob performed $\sigma_x$ on the second qubit of the Bell state, then he knows that Alice performed $\sigma_z$ on the first qubit of this Bell state. There will be four possibilities ($\sigma_z, \sigma_z$), ($\sigma_z, \sigma_x$), ($\sigma_x, \sigma_z$), ($\sigma_x, \sigma_x$) which can be encoded respectively into one bit classical information "0", "0", "1", "1", correspondingly and will be treated as raw key bits.

**Step.6** Alice and Bob will choose half of the key bits and disclose the values to check the eavesdroppers. If the error rate is higher than a predefined threshold, Alice and Bob will abort the protocol. Otherwise, they perform the privacy amplification [31,32] on the remained raw key bits to obtain the session key.

## 3. Security analyses

In this section, we will show the robustness of the proposed protocol. Robustness is introduced first by Boyer et al. [15,16] and means that an eavesdropper cannot obtain any information about the raw key without risking detection. In terms of security analysis, collective attack is a particularly important class of attacks because it includes most of well-known attacks, such as measurement attack, intercept-resend attack and etc. Furthermore, collective attack is the strongest joint attack (the most general attack) [33]. To prove the proposed protocol is robust, we have to prove that proposed protocol can be secure against the collective attack.

**Collective attack**

With the collective attack, the attacker will perform specific unitary operators to entangle an ancillary particle for each qubit. Then, the attacker measures the ancillary qubits to obtain useful information. We assume that TP is the attacker (Eve) because TP has more authority than other eavesdroppers in the protocol. If the proposed protocol can resist collective attack from the malicious TP, it can also resist the attack from other eavesdroppers. In this section, we want to prove that the attacker cannot obtain useful information without being detected. That is, if the attacker wants to obtain useful information, the attacker will introduce a detectable interruption to the original system and the attack will be detected.

**Theorem 1:** The proposed LMQKD is robust under collective attack.

**Proof:** We assume that TP generates the ancillary qubits $E = \{|E_1\rangle, |E_2\rangle, ...\}$ and performs a unitary operation $U_1$ on each qubit in Step 1. After Alice (Bob) finishing the operations in Step 2, TP performs another unitary operation $U_2$ on each qubit sent from Alice and Bob. Then TP can measure the ancillary qubits to obtain the measurement results and deduce the values of the shared key bits with the information of the ancillary qubits. First, we define $U_1$ operation and $U_2$ operation as follow:

$$U_1|\varphi^+\rangle_{AB} \otimes |E_1\rangle_E = a_1|\varphi^+\rangle|e_1\rangle + a_2|\varphi^-\rangle|e_2\rangle + a_3|\psi^+\rangle|e_3\rangle + a_4|\psi^-\rangle|e_4\rangle \quad \text{Eq. (1)}$$

where $|E_1\rangle$ denotes the initial state of ancillary qubit; $|a_1|^2 + |a_2|^2 + |a_3|^2 + |a_4|^2 = 1$; $|e_1\rangle$, $|e_2\rangle$, $|e_3\rangle$ and $|e_4\rangle$ are four states which are distinguishable to TP.

$$\begin{aligned}
U_2|\varphi^+\rangle_{AB} \otimes |E_2\rangle_E &= A_1|\varphi^+\rangle|F_1\rangle + A_2|\varphi^-\rangle|F_2\rangle + A_3|\psi^+\rangle|F_3\rangle + A_4|\psi^-\rangle|F_4\rangle \\
U_2|\varphi^-\rangle_{AB} \otimes |E_2\rangle_E &= B_1|\varphi^+\rangle|G_1\rangle + B_2|\varphi^-\rangle|G_2\rangle + B_3|\psi^+\rangle|G_3\rangle + B_4|\psi^-\rangle|G_4\rangle \\
U_2|\psi^+\rangle_{AB} \otimes |E_2\rangle_E &= C_1|\varphi^+\rangle|H_1\rangle + C_2|\varphi^-\rangle|H_2\rangle + C_3|\psi^+\rangle|H_3\rangle + C_4|\psi^-\rangle|H_4\rangle \\
U_2|\psi^-\rangle_{AB} \otimes |E_2\rangle_E &= D_1|\varphi^+\rangle|K_1\rangle + D_2|\varphi^-\rangle|K_2\rangle + D_3|\psi^+\rangle|K_3\rangle + D_4|\psi^-\rangle|K_4\rangle
\end{aligned} \quad \text{Eq. (2)}$$

where $|E_2\rangle$ also denotes the initial state of ancillary qubit; $|F_1\rangle$, $|F_2\rangle$, $|F_3\rangle$, $|F_4\rangle$, $|G_1\rangle$, $|G_2\rangle$, $|G_3\rangle$, $|G_4\rangle$, $|H_1\rangle$, $|H_2\rangle$, $|H_3\rangle$, $|H_4\rangle$, $|K_1\rangle$, $|K_2\rangle$, $|K_3\rangle$, $|K_4\rangle$

are sixteen states which are distinguishable to TP; $|A_1|^2+|A_2|^2+|A_3|^2+|A_4|^2=1$,

$|B_1|^2+|B_2|^2+|B_3|^2+|B_4|^2=1$, $|C_1|^2+|C_2|^2+|C_3|^2+|C_4|^2=1$ and

$|D_1|^2+|D_2|^2+|D_3|^2+|D_4|^2=1$ .

After Alice's and Bob's operations, TP computes all possible equations in **Group 1** and **Group 2** (Step 5) by Eq. (1) and Eq. (2).

Alice performs $\sigma_z$ and Bob performs $\sigma_z$:

$$U_2(a_1|\varphi^+\rangle|e_1\rangle - a_2|\varphi^-\rangle|e_2\rangle + a_3|\psi^+\rangle|e_3\rangle - a_4|\psi^-\rangle|e_4\rangle) \otimes |E_2\rangle_E$$
$$= a_1(A_1|\varphi^+\rangle|F_1\rangle + A_2|\varphi^-\rangle|F_2\rangle + A_3|\psi^+\rangle|F_3\rangle + A_4|\psi^-\rangle|F_4\rangle)$$
$$- a_2(B_1|\varphi^+\rangle|G_1\rangle + B_2|\varphi^-\rangle|G_2\rangle + B_3|\psi^+\rangle|G_3\rangle + B_4|\psi^-\rangle|G_4\rangle)$$
$$+ a_3(C_1|\varphi^+\rangle|H_1\rangle + C_2|\varphi^-\rangle|H_2\rangle + C_3|\psi^+\rangle|H_3\rangle + C_4|\psi^-\rangle|H_4\rangle) \quad Eq.\ (3)$$
$$- a_4(D_1|\varphi^+\rangle|K_1\rangle + D_2|\varphi^-\rangle|K_2\rangle + D_3|\psi^+\rangle|K_3\rangle + D_4|\psi^-\rangle|K_4\rangle)$$

Alice performs $\sigma_z$ and Bob performs $\sigma_x$:

$$U_2(a_1|\psi^-\rangle|e_1\rangle + a_2|\psi^+\rangle|e_2\rangle + a_3|\varphi^-\rangle|e_3\rangle + a_4|\varphi^+\rangle|e_4\rangle) \otimes |E_2\rangle_E$$
$$= a_4(A_1|\varphi^+\rangle|F_1\rangle + A_2|\varphi^-\rangle|F_2\rangle + A_3|\psi^+\rangle|F_3\rangle + A_4|\psi^-\rangle|F_4\rangle)$$
$$+ a_3(B_1|\varphi^+\rangle|G_1\rangle + B_2|\varphi^-\rangle|G_2\rangle + B_3|\psi^+\rangle|G_3\rangle + B_4|\psi^-\rangle|G_4\rangle)$$
$$+ a_2(C_1|\varphi^+\rangle|H_1\rangle + C_2|\varphi^-\rangle|H_2\rangle + C_3|\psi^+\rangle|H_3\rangle + C_4|\psi^-\rangle|H_4\rangle) \quad Eq.\ (4)$$
$$+ a_1(D_1|\varphi^+\rangle|K_1\rangle + D_2|\varphi^-\rangle|K_2\rangle + D_3|\psi^+\rangle|K_3\rangle + D_4|\psi^-\rangle|K_4\rangle)$$

Alice performs $\sigma_x$ and Bob performs $\sigma_z$:

$$U_2(-a_1|\psi^-\rangle|e_1\rangle + a_2|\psi^+\rangle|e_2\rangle + a_3|\varphi^-\rangle|e_3\rangle - a_4|\varphi^+\rangle|e_4\rangle) \otimes |E_2\rangle_E$$
$$= -a_4(A_1|\varphi^+\rangle|F_1\rangle + A_2|\varphi^-\rangle|F_2\rangle + A_3|\psi^+\rangle|F_3\rangle + A_4|\psi^-\rangle|F_4\rangle)$$
$$+ a_3(B_1|\varphi^+\rangle|G_1\rangle + B_2|\varphi^-\rangle|G_2\rangle + B_3|\psi^+\rangle|G_3\rangle + B_4|\psi^-\rangle|G_4\rangle)$$
$$+ a_2(C_1|\varphi^+\rangle|H_1\rangle + C_2|\varphi^-\rangle|H_2\rangle + C_3|\psi^+\rangle|H_3\rangle + C_4|\psi^-\rangle|H_4\rangle) \quad Eq.\ (5)$$
$$- a_1(D_1|\varphi^+\rangle|K_1\rangle + D_2|\varphi^-\rangle|K_2\rangle + D_3|\psi^+\rangle|K_3\rangle + D_4|\psi^-\rangle|K_4\rangle)$$

Alice performs $\sigma_x$ and Bob performs $\sigma_x$:

$$U_2(a_1|\varphi^+\rangle|e_1\rangle - a_2|\varphi^-\rangle|e_2\rangle + a_3|\psi^+\rangle|e_3\rangle - a_4|\psi^-\rangle|e_4\rangle)\otimes|E_2\rangle_E$$
$$= a_1(A_1|\varphi^+\rangle|F_1\rangle + A_2|\varphi^-\rangle|F_2\rangle + A_3|\psi^+\rangle|F_3\rangle + A_4|\psi^-\rangle|F_4\rangle)$$
$$- a_2(B_1|\varphi^+\rangle|G_1\rangle + B_2|\varphi^-\rangle|G_2\rangle + B_3|\psi^+\rangle|G_3\rangle + B_4|\psi^-\rangle|G_4\rangle)$$
$$+ a_3(C_1|\varphi^+\rangle|H_1\rangle + C_2|\varphi^-\rangle|H_2\rangle + C_3|\psi^+\rangle|H_3\rangle + C_4|\psi^-\rangle|H_4\rangle) \quad Eq.\ (6)$$
$$- a_4(D_1|\varphi^+\rangle|K_1\rangle + D_2|\varphi^-\rangle|K_2\rangle + D_3|\psi^+\rangle|K_3\rangle + D_4|\psi^-\rangle|K_4\rangle)$$

Now, TP has all possible equations after $U_2$ operation in key case (**Group 2**), then TP needs to compute the equation in check case (**Group 1**) to get the limitation of ancillary qubits (Note that the limitation from the **Group 1** are the same, so in our case we only take one scenario for illustration ).

Alice performs $H$ and Bob performs $H$:

$$U_2(a_1|\varphi^+\rangle|e_1\rangle + a_2|\psi^+\rangle|e_2\rangle + a_3|\varphi^-\rangle|e_3\rangle - a_4|\psi^-\rangle|e_4\rangle)\otimes|E_2\rangle_E$$
$$= a_1(A_1|\varphi^+\rangle|F_1\rangle + A_2|\varphi^-\rangle|F_2\rangle + A_3|\psi^+\rangle|F_3\rangle + A_4|\psi^-\rangle|F_4\rangle)$$
$$+ a_2(B_1|\varphi^+\rangle|G_1\rangle + B_2|\varphi^-\rangle|G_2\rangle + B_3|\psi^+\rangle|G_3\rangle + B_4|\psi^-\rangle|G_4\rangle)$$
$$+ a_3(C_1|\varphi^+\rangle|H_1\rangle + C_2|\varphi^-\rangle|H_2\rangle + C_3|\psi^+\rangle|H_3\rangle + C_4|\psi^-\rangle|H_4\rangle) \quad Eq.\ (7)$$
$$- a_4(D_1|\varphi^+\rangle|K_1\rangle + D_2|\varphi^-\rangle|K_2\rangle + D_3|\psi^+\rangle|K_3\rangle + D_4|\psi^-\rangle|K_4\rangle)$$

From Eq. (7), TP must set

$$a_1A_2|F_2\rangle + a_2B_2|G_2\rangle + a_3C_2|H_2\rangle - a_4D_2|K_2\rangle = a_1A_3|F_3\rangle + a_2B_3|G_3\rangle + a_3C_3|H_3\rangle - a_4D_3|K_3\rangle$$

$$= a_1A_4|F_4\rangle + a_2B_4|G_4\rangle + a_3C_4|H_4\rangle - a_4D_4|K_4\rangle = 0$$ to pass the Alice and Bob's detection. Then TP can get the condition: $A_2|F_2\rangle = D_2|K_2\rangle$, $A_3|F_3\rangle = D_3|K_3\rangle$, $A_4|F_4\rangle = D_4|K_4\rangle$, $B_2|G_2\rangle = B_3|G_3\rangle = B_4|G_4\rangle$ and $C_2|H_2\rangle = C_3|H_3\rangle = C_4|H_4\rangle$. Next, TP would substitute the deduced limitation in Eq. (3-6) to get the following equations:

$$a_1(A_1|\varphi^+\rangle|F_1\rangle + A_4|\psi^+\rangle|F_4\rangle)$$
$$+ a_2(B_2|\varphi^+\rangle|G_2\rangle + B_3|\psi^+\rangle|G_3\rangle)$$
$$+ a_3(C_2|\varphi^+\rangle|H_2\rangle + C_3|\psi^+\rangle|H_3\rangle) \quad Eq.\ (9)$$
$$+ a_4(D_1|\varphi^+\rangle|K_1\rangle + D_4|\psi^+\rangle|K_4\rangle)$$

$$a_2(A_1|\varphi^+\rangle|F_1\rangle + A_4|\psi^+\rangle|F_4\rangle)$$
$$+ a_1(B_2|\varphi^+\rangle|G_2\rangle + B_3|\psi^+\rangle|G_3\rangle)$$
$$+ a_3(C_2|\varphi^+\rangle|H_2\rangle + C_3|\psi^+\rangle|H_3\rangle)$$
$$+ a_4(D_1|\varphi^+\rangle|K_1\rangle + D_4|\psi^+\rangle|K_4\rangle)$$
Eq. (10)

$$a_3(A_1|\varphi^+\rangle|F_1\rangle + A_4|\psi^+\rangle|F_4\rangle)$$
$$+ a_4(B_2|\varphi^+\rangle|G_2\rangle + B_3|\psi^+\rangle|G_3\rangle)$$
$$+ a_1(C_2|\varphi^+\rangle|H_2\rangle + C_3|\psi^+\rangle|H_3\rangle)$$
$$+ a_2(D_1|\varphi^+\rangle|K_1\rangle + D_4|\psi^+\rangle|K_4\rangle)$$
Eq. (11)

$$a_4(A_1|\varphi^+\rangle|F_1\rangle + A_4|\psi^+\rangle|F_4\rangle)$$
$$+ a_3(B_2|\varphi^+\rangle|G_2\rangle + B_3|\psi^+\rangle|G_3\rangle)$$
$$+ a_2(C_2|\varphi^+\rangle|H_2\rangle + C_3|\psi^+\rangle|H_3\rangle)$$
$$+ a_1(D_1|\varphi^+\rangle|K_1\rangle + D_4|\psi^+\rangle|K_4\rangle)$$
Eq. (12)

According to Eq. (9-12), because $A_2|F_2\rangle = D_2|K_2\rangle$, $A_3|F_3\rangle = D_3|K_3\rangle$, $A_4|F_4\rangle = D_4|K_4\rangle$, TP could obtain the information about whether Alice's and Bob's operations are the same or not. This information unfortunately is not useful in revealing the key bits in our protocol. In other words, though TP can perform the unitary operation $U_1$ and $U_2$ (the operations that comply with the quantum mechanical theorems) to attack all transmitted qubits, there is no unitary operation for TP to obtain information about the participants' secret key without being detected.

## 4. Comparisons

This subsection compares the proposed MQKD protocol to the other existing MQKD protocols which are also three-party QKD with untrusted TP (see also **Table 2**).

In Shih's protocol [34,35], it allows the participants to perform three quantum operations including reordering the qubits. However, in the proposed protocol, it allows the participants to perform only two simple quantum operations. It is obvious that the proposed protocol is more lightweight in terms of participants' burden on quantum operations than theirs. Furthermore, in Krawec's and Liu's protocols, the participants have to reflect, generate, measure and reorder the single photons. These quantum operations are more difficult in implementation than performing lightweight quantum unitary operations according to the order of lightweight quantum operations discussed in Sec. 1. As for the qubit efficiency which is defined by the following formula: $QE = \frac{b_s}{q_t}$, where the parameter $b_s$ denotes the number of shared key bits in the end and the parameter $q_t$ denotes the number of the total particles used in the protocol. For $2m$ shared session key bits, the proposed protocol need to generate $9m$ pairs of Bell States, $m \in Z^+$. The qubit efficiency of the proposed protocol is $\frac{1}{9}$. On the other hand, and the qubit efficiencies of Shih's scheme [34-35], Krawec's scheme [26] and Liu's scheme [27] are $\frac{1}{2}$, $\frac{1}{24}$ and $\frac{1}{8}$, respectively.

**Table 2.** Comparison with other existing mediated QKD protocols

|  | Improved Shih's [34,35] | Krawec's [26] | Liu's [27] | Proposed protocol |
|---|---|---|---|---|
| **Protocol type** | Ordinary quantum | Semi-quantum | Semi-quantum | Lightweight quantum |
| **TP's quantum capabilities** | 1. Perform Z-basis measurement 2. Prepare the single photon in X basis | 1. Perform Bell measurement 2. Prepare the Bell states | 1. Perform Bell measurement 2. Prepare the Bell states | 1. Perform Bell measurement 2. Prepare the Bell states |
| **Participant's quantum capabilities** | 1. Unitary operations 2. Reflect 3. Reorder | 1. Measure 2. Prepare 3. Reflect | 1. Prepare 2. Reflect 3. Reorder | 1. Unitary operations 2. Reflect |
| **Qubit resource** | Single photons | Bell states | Single photons Bell states | Bell states |
| **Robustness Security Proof** | No | Yes | No | Yes |
| **Qubit efficiency** | $\frac{1}{2}$ | $\frac{1}{24}$ | $\frac{1}{8}$ | $\frac{1}{9}$ |

## 5. Conclusions

This study defines a new lightweight quantum environment allowing lightweight participants with only very lightweight quantum operations to perform quantum security protocols. In particular, a Lightweight Mediated QKD (LMQKD) which allows the lightweight quantum users to perform only: (1) unitary operations and (2) reflecting qubits is proposed. The qubit efficiency of the proposed protocol is better than Krawec's

SQKD protocol, and the participants' quantum capabilities are even more lightweight than Bob's in BB84 protocol and other SQKD protocols. That is, the proposed LMQKD protocol is practical in terms of the protocol implementation. Furthermore, the security analysis has shown that the proposed LMQKD protocol is robust under the collective attacks. In addition, according to survey [36-53], it is worth noting that the proposed definition of the lightweight quantum protocols covers some of the Measurement Device Independent (MDI) protocols. In terms of implementing the protocol, the proposed protocol can be practiced by the similar method used in [44]. It indeed is interesting in the future to design other lightweight quantum security protocols for various applications.

**Acknowledgment**

This research is partially supported by the Ministry of Science and Technology, Taiwan, R.O.C., under the Contract No. MOST 108-2221-E-006-107-; MOST 108-2627-E-006-001 -.